\documentclass[aps, prd, reprint, longbibliography, nofootinbib,superscriptaddress, floatfix]{revtex4-2}
\pdfoutput=1

\usepackage{amsmath,amssymb,amsfonts}
\usepackage{mathrsfs}
\usepackage{bbm}
\usepackage{slashed}
\usepackage{graphicx}
\usepackage{comment}
\usepackage{verbatim}
\usepackage[T1]{fontenc}
\usepackage[utf8]{inputenc}
\usepackage[colorlinks]{hyperref}
\usepackage{mathbbol}
\usepackage[dvipsnames,table]{xcolor}
\usepackage[normalem]{ulem}
\usepackage{svg}
\usepackage{flushend}
\usepackage{physics}
\usepackage{multirow}
\usepackage{graphicx}
\usepackage{dcolumn}
\usepackage{bm}
\usepackage{tensor}
\usepackage{comment}
\usepackage[utf8]{inputenc} 
\usepackage{amsthm,amsmath,amssymb,hyperref,mathrsfs}
\usepackage{braket,bm,bbm}
\usepackage{cancel}
\PassOptionsToPackage{normalem}{ulem}
\usepackage{ulem} 
\usepackage{physics}
\usepackage{float}
\usepackage[english]{babel}
\usepackage{graphicx}
\hypersetup{
    colorlinks=false,
    pdfborder={0 0 0},
}
\usepackage{comment}

\begin{document} 

\title{Impact of quantum gravity on the UV sensitivity of extremal black holes}

\author{Francesco Del Porro}
\email[]{francesco.del.porro@nbi.ku.dk}
\affiliation{Center of Gravity, Niels Bohr Institute, Blegdamsvej 17, DK-2100 Copenhagen \O, Denmark}
\affiliation{Niels Bohr International Academy, Niels Bohr Institute, Blegdamsvej 17, DK-2100 Copenhagen \O, Denmark}

\author{Francesco Ferrarin}
\email[]{francesco.ferrarin@nbi.ku.dk}
\affiliation{Center of Gravity, Niels Bohr Institute, Blegdamsvej 17, DK-2100 Copenhagen \O, Denmark}
\affiliation{Niels Bohr International Academy, Niels Bohr Institute, Blegdamsvej 17, DK-2100 Copenhagen \O, Denmark}

\author{Alessia Platania}
\email[]{alessia.platania@nbi.ku.dk}
\affiliation{Center of Gravity, Niels Bohr Institute, Blegdamsvej 17, DK-2100 Copenhagen \O, Denmark}
\affiliation{Niels Bohr International Academy, Niels Bohr Institute, Blegdamsvej 17, DK-2100 Copenhagen \O, Denmark}

\begin{abstract}
    Recent work has revealed that extremal Kerr black holes may exhibit a sensitivity to higher-derivative corrections to Einstein's equations, displaying singularities in the tidal forces at the horizon. However, in a purely gravitational context, this ``ultraviolet sensitivity'' translates into a strong dependence on the Wilson coefficients in the low-energy effective field theory. These, in turn, are fixed by the underlying theory of quantum gravity in the ultraviolet. We find a prediction for these coefficients within the framework of asymptotically safe quantum gravity, and show that, if the quantum gravity scale is trans-Planckian, this horizon-scale ultraviolet sensitivity is avoided.
\end{abstract}

\maketitle

\section{Introduction}\label{sect:introduction}

Effective field theory (EFT) corrections to general relativity (GR) are expected on general grounds and ought to be suppressed by appropriate powers of the EFT cutoff, making them more and more irrelevant at large distance scales. This view has been recently challenged: such corrections appear to be enhanced at the horizon of extremal Kerr geometries, making tidal forces divergent~\cite{Horowitz:2023xyl}. 
These divergences have been found not only to persist, but even to be stronger, in the case of extremal Kerr-Newman black holes~\cite{Horowitz:2024dch}. Such a behavior may be associated with a breakdown of EFT, despite its expected validity: although curvature invariants remain finite at the horizon, tidal forces grow unboundedly, making horizon-scale physics uniquely sensitive to ultraviolet (UV) physics, regardless of the black hole's mass.

Crucially, in a purely gravitational context, the emergence of these singularities is not universal; it is highly dependent on the values of the Wilson coefficients of the EFT that capture the low-energy gravitational dynamics. This dependence connects to the question of whether (specific theories of) quantum gravity (QG)~\cite{Bambi:2023jiz,Basile:2024oms,Buoninfante:2024yth} could prevent horizon-scale physics from developing UV sensitivity. Each UV completion of gravity is expected to map onto a landscape of low-energy EFTs that can be parameterized by the corresponding Wilson coefficients. Two theories in which the connection between UV physics and EFT is apparent are string theory (ST) and asymptotically safe quantum gravity (ASQG). In particular, in quantum field theory (QFT)-based formulations of QG, the construction of the QG landscape is well-defined and a clear recipe exists~\cite{Basile:2021krr,Knorr:2024yiu}.

Among the field theoretic approaches, ASQG~\cite{Knorr:2022dsx, Eichhorn:2022gku, Morris:2022btf, Martini:2022sll, Wetterich:2022ncl, Platania:2023srt, Saueressig:2023irs, Pawlowski:2023gym, Bonanno:2024xne} is particularly appealing: rooted in the Wilsonian renormalization group (RG), ASQG describes gravity as a consistent QFT at every scale, due to the existence of a non-Gaussian fixed point (NGFP) in the RG flow, which serves as an interacting UV completion of the theory. This generalizes the concept of asymptotic freedom, where the UV completion is a free theory, as in Quantum Chromodynamics. The existence of this so-called Reuter fixed point~\cite{Reuter:1996cp} has been corroborated within a massive amount of different approximations~\cite{Reuter:2001ag, Lauscher:2002sq, Codello:2006in, Gies:2015tca, Gies:2016con,Denz:2016qks, Hamada:2017rvn, Knorr:2017fus, Christiansen:2017bsy, Falls:2017lst, Falls:2020qhj, Knorr:2021slg, Kluth:2022vnq, Baldazzi:2023pep}, also including matter~\cite{Dona:2013qba,Christiansen:2017cxa,Pastor-Gutierrez:2022nki}, making ASQG a viable candidate for a non-perturbatively renormalizable and predictive theory of QG. If the resulting UV critical surface is finite-dimensional, as consistently indicated by RG computations~\cite{DeBrito:2018hur, Gies:2016con, Falls:2017lst, Kluth:2020bdv, Falls:2020qhj}, then the low-energy EFT is parametrized by a finite number of free couplings. While the conceptual question of black hole thermodynamics~\cite{Shomer:2007vq,Platania:2025imw,Basile:2025zjc} remains open, the coupling of ASQG with matter has showcased its predictive power, indicating that, if the gravitational RG flow departs from (or close to) the Reuter fixed point, the top-to-Higgs mass ratio and other standard model parameters turn out to be compatible with their center experimental values~\cite{Shaposhnikov:2009pv, Eichhorn:2017ylw, Eichhorn:2017lry, Eichhorn:2018whv, Eichhorn:2022gku,Eichhorn:2025sux}. 
These successes suggest that even if ASQG is not fundamental, the Reuter fixed point could act as a pivot, connecting the EFT regime to a more fundamental description like ST. In this scenario -- known as effective asymptotic safety~\cite{Percacci:2010af, deAlwis:2019aud, Held:2020kze} -- the predictive power of the Reuter fixed point remains unaffected, provided that in the QFT regime the RG trajectories stemming from the fundamental theory pass close to it. Hence, while the computation of Wilson coefficients from ASQG identifies specifically an ``asymptotic safety landscape'', in a scenario of effective ASQG, their specific values may not differ substantially from those of an underlying consistent fundamental theory.

In this letter, we focus on ASQG and tackle the question of whether, within this scenario, horizon-scale physics displays divergences in the tidal forces of extremal rotating black holes. Considering a truncation of the effective action containing up to six-derivative corrections to GR, we compute the asymptotic safety landscape and show that these singularities do not appear, provided that the scale of QG is at least Planckian. \vfill

\section{Background and setup}\label{sect:setup}

In this section, we briefly revise the results of~\cite{Horowitz:2023xyl}, where it is shown that EFT corrections can make horizon-scale tidal forces divergent. We then use these findings as a starting point to discuss RG flows in this setting and set up our calculation in the context of ASQG.

\subsection{EFT-corrected Extremal Kerr Black Holes}

It is expected that QG fluctuations, independently of the specific UV completion of gravity~\cite{Bambi:2023jiz, Basile:2024oms, Buoninfante:2024yth}, will complement the Einstein-Hilbert Lagrangian with an infinite tower of corrections, which can be organized in the form of an EFT expansion. The analysis of~\cite{Horowitz:2023xyl} focuses on a purely gravitational EFT, including essential operators up to eight-derivative terms. In four dimensions, the Gauss-Bonnet operator is a topological invariant and thus does not contribute to the equations of motion. As a result, the EFT Lagrangian in Lorentzian signature reads
\begin{equation}
    \mathcal{L}  =  \frac{1}{2 \kappa^2}  \left ( R  + \eta \kappa^4  \mathcal{R}^3 + \lambda \kappa^6  \mathcal{K}^2  +  \tilde \lambda  \kappa^6  \tilde{\mathcal{K}}^2 \right ) \, , \label{eq:Horowitz_Lagrangian}
\end{equation}
where $\kappa^2=8\pi G$, $(\eta,\lambda,\tilde{\lambda})$ are Wilson coefficients, and the higher-curvature terms are $ \mathcal{R}^3 \equiv R_{\mu \nu}\,^{\rho \sigma} R_{\rho \sigma}\,^{\lambda \tau} R_{\lambda \tau}\,^{\mu \nu} $, $ \mathcal{K} \equiv R^{\mu \nu \rho \sigma} R_{\mu \nu \rho \sigma}$, and $ \tilde{\mathcal{K}} \equiv R^{\mu \nu \rho \sigma} \tilde{R}_{\mu \nu \rho \sigma}$, with~$ \tilde{R}_{\mu \nu \rho \sigma} \equiv  \epsilon_{\mu \nu}\,^{\lambda \tau} R_{\lambda \tau \rho \sigma}$ being the dual Riemann tensor. We remark, at this point, that~\eqref{eq:Horowitz_Lagrangian} is a local EFT, where logarithmic form factors are neglected~\cite{Donoghue:2015xla,Donoghue:2018izj,Donoghue:2019fcb}.

Including higher-derivative corrections amounts to a change in the equations of motion, leading to deviations from Einstein's equations. In particular, these modifications affect the geometry of maximally rotating black holes in a potentially dramatic way, as they might induce singularities in the tidal forces at the horizon. Specifically, if one considers time-independent -- linearized in $(\eta, \, \lambda, \, \tilde \lambda)$ -- metric perturbations on top of the Kerr background, some components of the Weyl tensor display a power-law behavior close to the horizon,
\begin{equation} \label{eq:Weyl_scaling}
    C_{\rho a \rho b} \propto \rho^{\gamma -2}, \quad \gamma = \gamma^{(0)} + \eta \gamma^{(6)} + \lambda \gamma^{(8)} + \tilde \lambda \tilde \gamma^{(8)} \, ,
\end{equation}
where $\rho$ stands for the affine distance from the horizon. This scaling is related to the $ \text{O}(2,1)  \times  \text{U}(1)$ symmetry of the near-horizon geometry. The leading-order exponent is $ \gamma^{(0)} = 2$, indicating that the extremal horizon is smooth for Einstein gravity, or, in physical terms, tidal forces remain finite. The key finding of~\cite{Horowitz:2023xyl} is that EFT corrections $\{ \gamma^{(6)}, \,\gamma^{(8)},\, \tilde \gamma^{(8)} \}$ perturb this critical exponent, shifting it away from $2$ and potentially leading to a divergence if $\gamma -2 <0$. This is surprising, as EFT corrections to Einstein gravity are expected to only yield sub-leading corrections to GR. Yet, in the case of extremal rotating black holes, regardless of their masses, horizon-scale physics seems to be sensitive to the specific higher-derivative corrections generated by a putative UV completion of gravity (and matter). This UV sensitivity raises the question of whether any UV completion of gravity could yield Wilson coefficients, making the tidal forces finite. To answer this question in the context of ASQG, in the following, we shall focus on dynamics including up to six derivative terms, in which case the presence of a divergence is solely contingent upon the sign of $\eta$: the tidal forces diverge if $\eta<0$. 

\subsection{RG flows up to six derivatives}

Assessing the positivity of the Wilson coefficient~$\eta$ in QFT-based approaches to QG boils down to computing its RG flow. To this end, a natural starting point is to consider the flow of the Goroff-Sagnotti two-loop counterterm~\cite{Goroff:1985sz, Goroff:1985th, vandeVen:1991gw}, $\mathcal{C}^3 \equiv C_{\tau \lambda}\,^{\mu \nu} C_{\mu \nu}\,^{\rho \sigma}C_{\rho \sigma}\,^{\tau \lambda}$. Indeed, as the latter is related to the operator $\mathcal{R}^3$ in Eq.~\eqref{eq:Horowitz_Lagrangian} by a field redefinition, it is the only leading-order essential operator correcting Einstein gravity in $D=4$. Moreover, its RG flow has already been derived~\cite{Gies:2016con,Baldazzi:2023pep} in the context of ASQG within the following ``truncation'' of the gravitational dynamics in Euclidean signature\footnote{See~\cite{Manrique:2011jc,Biemans:2016rvp,Biemans:2017zca,Knorr:2018fdu,Eichhorn:2019ybe,Bonanno:2021squ,Fehre:2021eob,Banerjee:2022xvi,DAngelo:2022vsh,DAngelo:2023tis,Pawlowski:2025etp,Kher:2025rve} for steps towards Lorentzian RG computations.}
\begin{equation}
    \mathcal{L}  =  \frac{1}{2 \kappa^2}  (-R+ 2\Lambda) + G_{\mathcal{C}^3}  \mathcal{C}^3 \, . \label{eq:Goroff-Sagnotti}
\end{equation}
Crucially, the action above admits a(n asymptotically safe) UV completion~\cite{Gies:2016con,Baldazzi:2023pep}. The calculation has been extended recently and the result corroborated, using a different scheme~\cite{Baldazzi:2021ydj,Baldazzi:2021orb} which consistently accounts for essential couplings only~\cite{Baldazzi:2023pep,Knorr:2023usb}.

The correction $\eta \mathcal{R}^3$ can be mapped onto the Weyl-cubed operator via local field redefinitions involving only the metric and its derivatives. Notably, since the coupling $\eta$ is essential, no terms proportional to $\eta \mathcal{R}^3$ can be generated by field redefinitions from lower-order terms. Instead, it is straightforward to check that the field redefinition, which allows to switch between the two operator-basis ($\mathcal C^3$ and $\mathcal R^3$) for the essential scheme, takes the following form:
\begin{align}
  g_{\mu \nu} \to g_{\mu \nu}+ \eta \sum_i a_i O^i_{\mu \nu}(g) \,,
\end{align}
where $\{O_{\mu \nu}^i\}$ is a basis for the four-derivative tensors that can be built out of $g_{\mu \nu}$ and its derivatives, and $a_i$ are pure numbers~\cite{Fulling:1992vm}. This redefinition, being linear in $\eta$, leaves the equations of motion unaltered up to order $O(\eta^2)$. As a result, the field equations are unaffected at the linear level in $\eta$~\cite{Percacci:2017fkn}, which is the case considered in~\cite{Horowitz:2023xyl}. This means that the $(\rho a \rho b)$-component of the Weyl tensor, Eq.~\eqref{eq:Weyl_scaling}, on the horizon displays the same potentially pathological behavior with $G_{\mathcal{C}^3}$. Specifically, it diverges at the horizon if $G_{\mathcal{C}^3}<0$ \footnote{The sign of the coefficient remains unchanged when Wick-rotating back to Lorentzian signature.}. 

In what follows, we focus on the truncation in Eq.~\eqref{eq:Goroff-Sagnotti} for the full operator expansion, employing the beta functions obtained in~\cite{Baldazzi:2023pep} for the case of asymptotically flat spacetimes, $\Lambda = 0$. Although the truncation does not include terms up to eight derivatives as in~\cite{Horowitz:2023xyl}, cf. Eq.~\eqref{eq:Horowitz_Lagrangian}, it remains the highest-order truncation with all essential couplings for which the RG flow has been studied. 

\section{Asymptotic Safety prediction for the Goroff-Sagnotti coupling}
\label{sect:computing_the_landscape}

In this section, we will compute the asymptotic safety landscape, \textit{i.e.}, the set of EFTs stemming from ASQG, in the truncated dynamics~\eqref{eq:Goroff-Sagnotti}. We will obtain a prediction for the Goroff-Sagnotti coupling $G_{\mathcal{C}^3}$ and discuss its implications for the UV sensitivity of EFT-corrected rotating black holes. The Functional Renormalization Group (FRG)~\cite{Dupuis:2020fhh} provides a natural recipe for this purpose~\cite{Basile:2021krr,Knorr:2024yiu}. By introducing an auxiliary infrared (IR) cutoff scale $k$, the FRG implements the Wilsonian shell-by-shell integration of fluctuating modes in the path integral by defining a $k$-dependent version of the effective action, dubbed the effective average action (EAA), which interpolates between the bare action in the UV (limit $k\to\infty$) and the standard effective action in the IR (physical limit $k\to 0$). At a given scale $k$, the EAA defines a ``partial'' effective action resulting from the integration of fluctuations with momenta $p^2>k^2$. As a result, every coupling acquires a dependence on $k$. If the flow admits a UV fixed point, the theory is renormalizable -- with the fixed point acting as its UV completion -- and the IR limit $k\to0$ well-defined. In particular, the $k \to 0$ limit of UV complete FRG trajectories defines the set of EFTs stemming from this UV completion. Hence, the Wilson coefficients are identified with the IR limits of these RG running couplings\footnote{This RG running is not to be confused with the physical running of couplings, see~\cite{Bonanno:2020bil,Buccio:2023lzo,Buccio:2024hys}. Within the FRG, the running with the RG scale $k$ keeps track of the integrated fluctuating modes in the path integral, whereas the physical running with physical momenta $p$ is encoded in form factors in the EAA~\cite{Knorr:2019atm,Draper:2020knh,Knorr:2021iwv,Knorr:2022dsx}.}. In this context, the number of free dimensionless Wilson coefficients is set by the number of relevant directions of the fixed point, modulo one fixing the unit scale. All other dimensionless Wilson coefficients are determined by the UV completion in terms of the relevant ones.

\begin{figure*}[t!]
    \centering
    \includegraphics[width=0.7\linewidth]{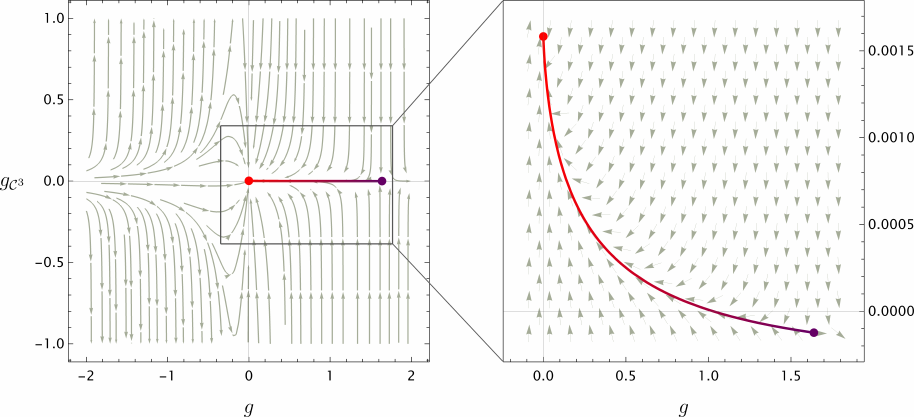}
    \caption{Different zooms on the RG flow resulting from the beta functions of~\cite{Baldazzi:2023pep}. The flow is plotted on the $\{g_{\mathcal{C}^3}(k),g(k)\}$ plane, and is a parametric plot of the running couplings. The arrows indicate their variations with $k$, pointing from the UV to the IR, $k\to0$. As apparent from the left panel, the structure of the flow is governed by two fixed points: an sGFP, which is IR attractive, and an NGFP, which is a saddle point, with one relevant and one irrelevant direction. There is thus a unique UV complete trajectory, depicted as a solid purple-to-red colored line, that is also the separatrix between the two fixed points; this is the trajectory to be used to extract the value of the Goroff-Sagnotti coupling. A zoomed-in version of the flow focusing on the separatrix is given in the right panel.} 
    \label{fig:RG_flow_shifted_GFP}
\end{figure*}

To extract physical predictions from ASQG, we need to study the flow of every possible FRG trajectory departing from the interacting fixed point -- the asymptotically safe UV completion of the theory. The RG flow is written in terms of the dimensionless couplings
\begin{equation*}
    g(k) = k^2 G_{\text{N}}(k), \qquad g_{\mathcal{C}^3}(k) = k^2 G_{\mathcal{C}^3}(k) \, .
\end{equation*}
Within the truncation employed here, and based on the beta functions of~\cite{Baldazzi:2023pep}, the flow admits two fixed points: a shifted Gaussian fixed point (sGFP) and an NGFP,
\begin{align*}
    &\text{sGFP:}  &&\quad g_0   =   0 \, ,         &&\quad g_{\mathcal{C}^3,0}   =   \frac{1}{64 \pi^2} \, ; \\
    &\text{NGFP:} &&\quad g_*   =   1.639 \, ,     &&\quad g_{\mathcal{C}^3,*}   =   -1.233 \cdot 10^{-4} \, .
\end{align*}
The interacting fixed point comes with critical exponents $(-12.88, 2.091)$, \textit{i.e.}, it presents a single relevant direction. As a consequence, there exists a unique UV complete RG trajectory, which is the separatrix connecting the NGFP in the UV to the sGFP in the IR: once the scale of QG is fixed, there exists a unique EFT in the asymptotic safety landscape stemming from the NGFP. We now want to identify this EFT by computing its Wilson coefficients. To this end, we first determine the separatrix by numerically integrating the beta functions, and initializing the flow near the NGFP with a perturbation along the relevant eigendirection. Fig.~\ref{fig:RG_flow_shifted_GFP} illustrates the full RG flow, highlighting the separatrix connecting the two fixed points.

Next, it is useful to find a fit for the UV complete trajectory in the IR. Since in the IR this RG trajectory ends at the sGFP, we can conveniently guess this fit by following the canonical mass-dimension scaling and accounting for the logarithmic running,
\begin{align}
        g_{\mathcal{C}^3}(k) &= g_{\mathcal{C}^3,0}  +g_{\mathcal{C}^3,\text{IR}}  \left ( \frac{k}{k_0} \right )^2  \left [ 1 + b\,\log\left( \frac{k}{k_0}\right)  \right] \, ,  \label{eq:alpha_scaling}\\
        g(k) &= g_{\text{IR}}  \left ( \frac{k}{k_0} \right)^2 \, , \label{eq:g_scaling}
\end{align}
where $k_0$ is an integration constant that sets the speed of the flow and has the interpretation of the scale of QG~\cite{Basile:2021krr, Knorr:2024yiu}, $g_{\text{IR}}$ is an arbitrary constant that sets the IR value of the Newton's coupling,   $g_{\mathcal{C}^3,\text{IR}}$ depends on the specific RG trajectory, and $b$ is the slope of the logarithm.

The IR logarithmic running of the Goroff-Sagnotti coupling is due to the integration of massless fluctuations. While such a feature is expected for marginal couplings, its appearance in the flow of $g_{\mathcal{C}^3}$ is non-trivially induced by the running of lower-order couplings, since they enter the higher-order beta functions. This logaritmic $k$-running is known to reflect the presence of physical logarithmic form factors $\log ( \Box/\Lambda^2_{\text{QG}})$ in the effective action, where $\Lambda_{\text{QG}}$ is a characteristic energy, which ought to set the transition scale to the full-fledged QG regime. This, in general, introduces an ambiguity in the definition of the Wilson coefficients~\cite{Alberte:2020jsk,Herrero-Valea:2020wxz,Herrero-Valea:2022lfd,Henriksson:2022oeu,Tokareva:2025rta}. Indeed, under a redefinition of the reference scale, $a + \log \small(k^2/k_0^2 \small) = \tilde{a} + \log \small(k^2/\tilde{k}_0^2 \small)$, the value of the Wilson coefficient (in our case $g_{\mathcal{C}^3}$) shifts accordingly. Hence, one needs either to perform the analysis of~\cite{Horowitz:2023xyl,Horowitz:2024dch} accounting for these logarithmic form factors, or to adopt a prescription to define the IR limit of $g_{\mathcal{C}^3}$.\footnote{Progress in investigating bounds on Wilson coefficients in the presence of massless fluctuations has been made in~\cite{Alberte:2020jsk,Herrero-Valea:2020wxz,Herrero-Valea:2022lfd,Henriksson:2022oeu,Tokareva:2025rta}.} We will proceed in the latter way, following the prescription in~\cite{Basile:2021krr, Knorr:2024yiu}, and identifying the transition scale with the Planck mass $M_{\text{Pl}}$. This is a natural choice in a fundamental realization of ASQG, since the Planck mass is the only relevant scale of the system. More concretely, the dimensionful Newton's constant is defined by the limit
\begin{equation}
    G_{\text{N}}  =  \lim_{k \to 0} \frac{ g(k) }{k^2}  =  \frac{g_{\text{IR}}}{k_0^2} \, ,
\end{equation}
which allows us to choose $g_{\text{IR}}  =  1$ and hence relate the transition scale to the Planck mass, $k_0^2  =  G_{\text{N}}^{-1}\equiv M_{\text{Pl}}^2$. It is in principle possible to choose another $g_{\text{IR}}\gg 1$ or $g_{\text{IR}}\ll 1$, implying a transition scale much below/above the Planck scale. However, this would be unnatural, since it would artificially introduce a new scale into the system. We will come back to this important point later.

Having identified the Planck mass to be the transition scale, we can now determine the Goroff-Sagnotti Wilson coefficient following the prescription in~\cite{Basile:2021krr,Knorr:2024yiu}, \textit{i.e.}, by noticing that Wilson coefficients in the chosen units can be computed as the IR limits of appropriate ratios of dimensionless couplings, and by subtracting the logarithmic running. In our case, once the logarithmic running is appropriately subtracted, the Goroff-Sagnotti Wilson coefficient in Planck units is defined by the limit 
\begin{equation}\label{eq:ratio}
     \frac{G_{\mathcal{C}^3}}{G_N}=\lim_{k\to 0} \frac{g_{\mathcal{C}^3}(k)}{g(k)}\,.
\end{equation}
However, the flow based on the beta functions in~\cite{Baldazzi:2023pep} presents an additional problem we need to deal with: since the GFP is shifted by a small amount, both ${G_{\mathcal{C}^3}}$ and the ratio above display unphysical IR nonlocalities of the $1/\Box$-type. This problem is generated by a regularization scheme that does not involve ``natural'' endomorphisms in the sense of~\cite{Knorr:2021slg} in all regulators. There are at least two ways to overcome this problem. A first possibility is to keep the same regularization scheme as in~\cite{Baldazzi:2023pep} and implement an \textit{ad hoc} subtraction of the unphysical divergence. Alternatively, one could employ a regularization scheme that does not lead to this issue, \textit{i.e.}, one where the IR physics is governed by a standard, non-shifted GFP. On general grounds, since the IR fixed point of the two cases is different, we may expect the Wilson coefficients to be different. However, since both flows would share the same UV physics, at least the qualitative outcome is expected to be the same -- provided that both procedures are correctly implemented. We will pursue both strategies independently to compare the results and test their robustness.

\textit{``Unnatural'' scheme and IR divergences subtraction ---} Due to the IR behavior of $g(k)$ and $g_{\mathcal{C}^3}(k)$, given by Eqs.~\eqref{eq:alpha_scaling} and~\eqref{eq:g_scaling}, we may define the Goroff-Sagnotti Wilson coefficient by considering the ratio of their deviations from their fixed point values,
\begin{equation}
    \frac{g_{\mathcal{C}^3}(k)-g_{\mathcal{C}^3,0}}{g(k)}   \, , 
    \label{eq:ratio_log_running}
\end{equation}
rather than the ratio in Eq.~\eqref{eq:ratio}. Intuitively, Eq.~\eqref{eq:ratio_log_running} represents the way in which the theory approaches its IR limit along the UV complete trajectory of the flow. This construction removes the unphysical IR divergence of the ${1}/{\Box}$-type, and isolates the true physical running.

As for the physical, expected logarithmic running, we eliminate it following the prescription outlined in~\cite{Basile:2021krr, Knorr:2024yiu},
\begin{equation}
    \frac{G_{\mathcal{C}^3}|_{k_0=M_{\text{Pl}}}}{G_{\text{N}}}  =  \frac{g_{\mathcal{C}^3}(k)-g_{\mathcal{C}^3,0}}{g(k)} -  g_{\mathcal{C}^3,\text{IR}}b \log(\frac{k}{M_{\text{Pl}}}) \, , \label{eq:scale_fixing}
\end{equation}
with $g_{\mathcal{C}^3,\text{IR}}b = -0.011$
obtained from matching the IR numerical flow with Eqs.~\eqref{eq:alpha_scaling} and~\eqref{eq:g_scaling}. With this prescription, we obtain a positive Wilson coefficient for $G_{\mathcal{C}^3}$:
\begin{equation}
    \frac{G_{\mathcal{C}^3}|_{k_0=M_{\text{Pl}}}}{G_{\text{N}}}  =  9.6 \cdot 10^{-3} >0 \, . \label{fig:eq:prediction_first_scheme}
\end{equation}
This implies that the tidal forces at the EFT-corrected extremal Kerr horizon remain finite. 

Next, we are going to test these findings by using a different regularization scheme that avoids the unphysical IR nonlocalities.

\textit{Natural regularization scheme ---} To investigate the robustness of our result, we repeat the analysis in a different scheme in which the GFP is not shifted and Eq.~\eqref{eq:ratio_log_running} reduces to the ratio of the running dimensionless couplings, as in Eq.~\eqref{eq:ratio}. As anticipated, this requires modifying the regularization scheme of~\cite{Baldazzi:2023pep}, in a way that all regulators use ``natural'' endomorphisms~\cite{Knorr:2021slg}\footnote{We thank B. Knorr for private correspondence on this issue and for providing a modified version of the beta functions in~\cite{Baldazzi:2023pep} (see attached notebook).}. The flow admits a standard GFP, and an NGFP at
\begin{align*}
    &\text{NGFP:} &&\quad g_* = 0.5890 \, ,     &&\quad g_{\mathcal{C}^3,*} = -3.242 \cdot 10^{-7} \, .
\end{align*}
The critical exponents associated with the NGFP are $(-7.750, 2.783)$: their signs and magnitude are compatible with those of the previous case, indicating that the different scheme only weakly affects UV physics. At the same time, a convergence of the numerical value of the critical exponents requires adding more curvature invariants~\cite{Falls:2014tra,Falls:2017lst,Kluth:2020bdv}. Such numerical precision is beyond the scope of our work, as only the signs of the critical exponents and of the Goroff-Sagnotti Wilson coefficient matter for our conclusions. 

The RG flow resulting from the new beta functions is very similar to the one we previously discussed (see Fig.~\ref{fig:RG_flow_shifted_GFP}), modulo the location of the IR fixed point, which is now non-shifted. Once again, the Goroff-Sagnotti coupling presents logarithmic IR running, and therefore the ratio $g_{\mathcal{C}^3}(k)/g(k)$ has a (physically meaningful) logarithmic behavior, where the slope of the logarithm now reads $g_{\mathcal{C}^3,\text{IR}}b = -3.07 \cdot 10^{-6}$. Fixing $k_0=M_{\text{Pl}}$ and subtracting the logarithmic running as before, we can extract a new prediction for the Goroff-Sagnotti Wilson coefficients, which reads 
\begin{equation}
    G_{\mathcal{C}^3}|_{k_0=M_{\text{Pl}}}=3.02 \cdot 10^{-6}G_{\text{N}} \,.
\end{equation} 
Unsurprisingly, since in the two regularization schemes IR physics is governed by different fixed points, one of which leads to unphysical divergences that must be subtracted by hand, the order of magnitude of the two resulting Wilson coefficients is different. While this issue is to be better investigated in the future to improve the robustness and stability of the FRG computations, the qualitative feature that matters for the purpose of this letter is the sign of the Wilson coefficient, which remains unaltered by the choice of regularization scheme. This indicates that in ASQG, UV sensitivity of EFT-corrected rotating black holes may be avoided. 

\section{QG scale and UV (in)sensitivity}
\label{sect:Transition_scale_scheme_dependence}

We showed how ASQG predicts the right sign of the Wilson coefficient to avoid UV sensitivity on the horizon. However, the logarithmic running introduces an ambiguity in the definition of the Wilson coefficient~\cite{Alberte:2020jsk,Herrero-Valea:2020wxz,Herrero-Valea:2022lfd,Tokareva:2025rta}, and adopting a prescription for its subtraction is tantamount to introducing a dependence on the characteristic QG scale $k_0$. In the previous section, we fixed $k_0$ to be the Planck scale, since this is the only scale naturally appearing in a fundamental realization of ASQG. In this section, we are going to challenge this choice and analyze how our conclusions for the UV sensitivity of extremal black holes in ASQG may change, as we change $k_0$.

More explicitly, letting $k_0 \equiv \xi M_{\text{Pl}}$, the IR limit of the dimension Goroff-Sagnotti coupling shifts as
\begin{equation}
    g_{\mathcal{C}^3}|_{k_0}  =  g_{\mathcal{C}^3}|_{M_{\text{Pl}}}+  g_{\mathcal{C}^3,\text{IR}}b \log(\xi) \, . \label{eq:reference_scale}
\end{equation}
Although in general the choice of the transition scale does not affect the sign and magnitude of the Wilson coefficients~\cite{Basile:2021krr, Knorr:2024yiu}, the equation above
reveals a potentially pathological situation: if $g_{\mathcal{C}^3,\text{IR}}b<0$, a sufficiently large $\xi$ renders the sum in Eq.~\eqref{eq:reference_scale} negative, thus flipping the sign of the Goroff-Sagnotti Wilson coefficient. Demanding regularity at the horizon is thus equivalent to setting a lower bound on~$\xi$: the divergence of tidal forces is avoided for $\xi\geq \exp (-g_{\mathcal{C}^3}|_{M_{\text{Pl}}}/g_{\mathcal{C}^3,\text{IR}}b)\simeq 0.42$, \textit{i.e.}, if the QG scale is $k_0\gtrsim 0.42 M_{\text{Pl}}$. Notably, a very similar result is obtained when using the second, ``natural'' regularization scheme discussed in the previous section, in which case the bound reads $\xi\gtrsim 0.37$. Based on both bounds, we discover that the UV sensitivity of extremal black holes is avoided if the scale of QG is at least Planckian,
\begin{equation}
    k_0\gtrsim \mathcal{O}(10^{-1})M_{\text{Pl}}\,.
\end{equation}

This is expected in a fundamental realization of ASQG, since in this case, the Planck mass is the only dimensionful scale of the system. On the contrary, if ASQG is realized at an effective level, \textit{e.g.}, as a low-energy realization of ST~\cite{deAlwis:2019aud,Basile:2021euh,Basile:2021krk}, then its characteristic transition scale would need to lie below the string scale; based on general arguments~\cite{Basile:2023blg, Castellano:2023aum, Basile:2024dqq, Bedroya:2024ubj, Aoufia:2024awo}, the latter ought to be sub-Planckian, implying horizon-scale divergences of tidal forces.

Notably, these results appear to resonate with other complementary findings in the literature: in $D>4$, and including the $U(1)$ sector, UV sensitivity of extremal Reissner-Nordstr\"om black holes is avoided if the Wilson coefficients violate (even slightly)~\cite{Horowitz:2023xyl,Chen:2024sgx} the weak gravity conjecture (WGC)~\cite{Arkani-Hamed:2006emk, Banks:2006mm, Harlow:2022ich} for black holes. Although the WGC is expected to hold on general grounds~\cite{Cheung:2018cwt, Hamada:2018dde, Urbano:2018kax, Heidenreich:2024dmr}, Planck-scale-suppressed violations of the WGC for black holes are possible~\cite{Alberte:2020jsk,Henriksson:2022oeu}. Evidence for such violations has been found in~\cite{Knorr:2024yiu} in a scenario of fundamental ASQG. Yet, if ASQG were realized at an effective level, the RG trajectory realized by nature would not emanate from the NGFP, it would rather lie close to it; then, the fundamental UV completion (such as ST) would presumably lead to Wilson coefficients that are \textit{(i)} close but not equal to those of an effective ASQG scenario, and \textit{(ii)} satisfying the WGC for black holes. In this case, tidal forces would again be divergent, as opposed to a scenario of fundamental ASQG, and compatibly with our findings.

\section{Conclusions}

EFT corrections to rotating extremal black holes may lead to pathological divergences of the tidal forces at the event horizon~\cite{Horowitz:2023xyl}. At least in a purely gravitational setting, this result depends on the Wilson coefficients of the low-energy EFT, which in turn encode information about the UV completion of gravity. In this letter, we explored the question of whether this ``UV sensitivity'' of extremal rotating black holes can be avoided in the ASQG.

We focused on a truncation that includes all essential operators up to six derivatives, with vanishing cosmological constant. Only two couplings are thus present: the Newton’s constant and the Goroff-Sagnotti coupling. As shown in~\cite{Horowitz:2023xyl}, a positive value for the latter ensures regularity of tidal forces near extremal rotating horizons. In this setting, ASQG is characterized by an interacting fixed point with a single relevant direction. Thus, once the physical unit is fixed, the theory has no remaining free parameters: there exists a unique RG trajectory connecting the UV fixed point to low-energy physics, and hence a unique EFT stemming from it. However, extracting the Wilson coefficient requires dealing with logarithmic corrections induced by graviton fluctuations. Subtracting these logarithmic form factors as proposed in~\cite{Basile:2021krr,Knorr:2024yiu}, and assuming the characteristic scale of the logarithms -- which is related to the scale of QG -- to be the Planck mass, we obtain a positive Goroff-Sagnotti Wilson coefficient.
We have explicitly verified that this result is robust against changes in the regularization scheme used to construct the flow. This provides evidence that a fundamental realization of ASQG predicts regular tidal forces near extremal rotating black holes.

The choice of the Planck scale as the QG scale is justified because, in a fundamental realization of ASQG, the Planck mass is the only naturally available scale of the system. However, if this scale is allowed to deviate from the Planck mass -- as may happen in an effective realization of ASQG~\cite{deAlwis:2019aud,Basile:2021euh,Basile:2021krk} -- then regularity of tidal forces imposes a lower bound on the QG scale, which must be at least Planckian. If the UV completion lies below this scale, and if a regime of effective ASQG is realized at intermediate energies, tidal divergences reappear.

Speculatively, this outcome resonates with the picture drawn by various results beyond pure gravity. First, according to~\cite{Horowitz:2023xyl}, the validity of the WGC for black holes implies that tidal divergences are unavoidable in $D\geq5$. While the WGC appears to hold in known string constructions~\cite{Harlow:2022ich}, it might be slightly violated in a fundamental realization of ASQG~\cite{Knorr:2024yiu}. This ``near-horizon negativity'' may be key to curing the tidal divergences, as conjectured in~\cite{Chen:2024sgx}. However, if ASQG emerges effectively from a more fundamental description like ST, then the WGC would be inherited, and the associated divergences would be present. Accordingly, we observe a qualitative distinction between fundamental and effective realizations of ASQG: while the former may ensure regularity of tidal forces, the latter generically does not. 

It is key to further comment on the case of UV sensitivity in the presence of matter, in particular a $U(1)$ gauge field, in $D=4$. In this case, the problem of UV sensitivity is even more severe, since divergences are stronger and the dependence on the Wilson coefficients is weaker~\cite{Horowitz:2024dch}. The extension of our arguments to that case, and the possibility of curing UV sensitivity with trans-Planckian physics, would then require a combined UV completion of gravity and matter, and hence, perhaps, some form of grand unified theories. It would be interesting to explore how the results of~\cite{Horowitz:2024dch} would extend to that case, and whether the resulting combined UV completion could prevent horizon-scale tidal forces from diverging.

Finally, our analysis relies on a specific treatment of logarithmic corrections. A natural extension would be to incorporate the logarithmic form factors directly into the effective action, thereby eliminating the ambiguity in the definition of the Wilson coefficient at its source. We leave this for future investigations.

\acknowledgments
The authors would like to thank I. Basile, C. Chen, B. Knorr, and G. Remmen for discussions and B. Knorr for comments on the manuscript. This research is supported by a research grant (VIL60819) from VILLUM FONDEN. The Center of Gravity is a Center of Excellence funded by the Danish National Research Foundation under grant No. 184.

\bibliographystyle{apsrev4-1}
\bibliography{references}

@article{Fehre:2021eob,
    author = "Fehre, Jannik and Litim, Daniel F. and Pawlowski, Jan M. and Reichert, Manuel",
    title = "{Lorentzian Quantum Gravity and the Graviton Spectral Function}",
    eprint = "2111.13232",
    archivePrefix = "arXiv",
    primaryClass = "hep-th",
    doi = "10.1103/PhysRevLett.130.081501",
    journal = "Phys. Rev. Lett.",
    volume = "130",
    number = "8",
    pages = "081501",
    year = "2023"
}

@article{Biemans:2017zca,
    author = "Biemans, Jorn and Platania, Alessia and Saueressig, Frank",
    title = "{Renormalization group fixed points of foliated gravity-matter systems}",
    eprint = "1702.06539",
    archivePrefix = "arXiv",
    primaryClass = "hep-th",
    doi = "10.1007/JHEP05(2017)093",
    journal = "JHEP",
    volume = "05",
    pages = "093",
    year = "2017"
}

@article{Pawlowski:2025etp,
    author = "Pawlowski, Jan M. and Reichert, Manuel and Wessely, Jonas",
    title = "{Self-consistent graviton spectral function in Lorentzian quantum gravity}",
    eprint = "2507.22169",
    archivePrefix = "arXiv",
    primaryClass = "hep-th",
    month = "7",
    year = "2025"
    }

@article{DAngelo:2022vsh,
    author = "D'Angelo, Edoardo and Drago, Nicol\`o and Pinamonti, Nicola and Rejzner, Kasia",
    title = "{An Algebraic QFT Approach to the Wetterich Equation on Lorentzian Manifolds}",
    eprint = "2202.07580",
    archivePrefix = "arXiv",
    primaryClass = "math-ph",
    doi = "10.1007/s00023-023-01348-4",
    journal = "Annales Henri Poincare",
    volume = "25",
    number = "4",
    pages = "2295--2352",
    year = "2024"
}

@unpublished{DAngelo:2023tis,
    author = "D'Angelo, Edoardo and Rejzner, Kasia",
    title = "{A Lorentzian renormalisation group equation for gauge theories}",
    eprint = "2303.01479",
    archivePrefix = "arXiv",
    primaryClass = "math-ph",
    month = "3",
    year = "2023",
    note = "{a}rXiv Preprint"
}

@article{Knorr:2018fdu,
    author = "Knorr, Benjamin",
    title = "{Lorentz symmetry is relevant}",
    eprint = "1810.07971",
    archivePrefix = "arXiv",
    primaryClass = "hep-th",
    doi = "10.1016/j.physletb.2019.01.070",
    journal = "Phys. Lett. B",
    volume = "792",
    pages = "142--148",
    year = "2019"
}

@article{Christiansen:2017bsy,
    author = "Christiansen, Nicolai and Falls, Kevin and Pawlowski, Jan M. and Reichert, Manuel",
    title = "{Curvature dependence of quantum gravity}",
    eprint = "1711.09259",
    archivePrefix = "arXiv",
    primaryClass = "hep-th",
    doi = "10.1103/PhysRevD.97.046007",
    journal = "Phys. Rev. D",
    volume = "97",
    number = "4",
    pages = "046007",
    year = "2018"
}

@article{DeBrito:2018hur,
    author = "De Brito, Gustavo P. and Ohta, Nobuyoshi and Pereira, Antonio D. and Tomaz, Anderson A. and Yamada, Masatoshi",
    title = "{Asymptotic safety and field parametrization dependence in the $f(R)$ truncation}",
    eprint = "1805.09656",
    archivePrefix = "arXiv",
    primaryClass = "hep-th",
    doi = "10.1103/PhysRevD.98.026027",
    journal = "Phys. Rev. D",
    volume = "98",
    number = "2",
    pages = "026027",
    year = "2018"
}

@article{Kluth:2022vnq,
    author = "Kluth, Yannick and Litim, Daniel F.",
    title = "{Functional renormalization for f(R{\ensuremath{\mu}}{\ensuremath{\nu}}{\ensuremath{\rho}}{\ensuremath{\sigma}}) quantum gravity}",
    eprint = "2202.10436",
    archivePrefix = "arXiv",
    primaryClass = "hep-th",
    doi = "10.1103/PhysRevD.106.106022",
    journal = "Phys. Rev. D",
    volume = "106",
    number = "10",
    pages = "106022",
    year = "2022"
}

@article{Christiansen:2017cxa,
    author = "Christiansen, Nicolai and Litim, Daniel F. and Pawlowski, Jan M. and Reichert, Manuel",
    title = "{Asymptotic safety of gravity with matter}",
    eprint = "1710.04669",
    archivePrefix = "arXiv",
    primaryClass = "hep-th",
    doi = "10.1103/PhysRevD.97.106012",
    journal = "Phys. Rev. D",
    volume = "97",
    number = "10",
    pages = "106012",
    year = "2018"
}

@article{Falls:2014tra,
    author = "Falls, Kevin and Litim, Daniel F. and Nikolakopoulos, Konstantinos and Rahmede, Christoph",
    title = "{Further evidence for asymptotic safety of quantum gravity}",
    eprint = "1410.4815",
    archivePrefix = "arXiv",
    primaryClass = "hep-th",
    reportNumber = "DO-TH-14-26, KA-TP-2014-30",
    doi = "10.1103/PhysRevD.93.104022",
    journal = "Phys. Rev. D",
    volume = "93",
    number = "10",
    pages = "104022",
    year = "2016"
}

@article{Knorr:2017fus,
    author = "Knorr, Benjamin and Lippoldt, Stefan",
    title = "{Correlation functions on a curved background}",
    eprint = "1707.01397",
    archivePrefix = "arXiv",
    primaryClass = "hep-th",
    doi = "10.1103/PhysRevD.96.065020",
    journal = "Phys. Rev. D",
    volume = "96",
    number = "6",
    pages = "065020",
    year = "2017"
}

@article{Banerjee:2022xvi,
    author = "Banerjee, Rudrajit and Niedermaier, Max",
    title = "{The spatial Functional Renormalization Group and Hadamard states on cosmological spacetimes}",
    eprint = "2201.02575",
    archivePrefix = "arXiv",
    primaryClass = "hep-th",
    doi = "10.1016/j.nuclphysb.2022.115814",
    journal = "Nucl. Phys. B",
    volume = "980",
    pages = "115814",
    year = "2022"
}

@article{Manrique:2011jc,
    author = "Manrique, Elisa and Rechenberger, Stefan and Saueressig, Frank",
    title = "{Asymptotically Safe Lorentzian Gravity}",
    eprint = "1102.5012",
    archivePrefix = "arXiv",
    primaryClass = "hep-th",
    reportNumber = "MZ-TH-11-02",
    doi = "10.1103/PhysRevLett.106.251302",
    journal = "Phys. Rev. Lett.",
    volume = "106",
    pages = "251302",
    year = "2011"
}

@article{Biemans:2016rvp,
    author = "Biemans, Jorn and Platania, Alessia and Saueressig, Frank",
    title = "{Quantum gravity on foliated spacetimes: Asymptotically safe and sound}",
    eprint = "1609.04813",
    archivePrefix = "arXiv",
    primaryClass = "hep-th",
    doi = "10.1103/PhysRevD.95.086013",
    journal = "Phys. Rev. D",
    volume = "95",
    number = "8",
    pages = "086013",
    year = "2017"
}

@article{Bonanno:2021squ,
    author = "Bonanno, Alfio and Denz, Tobias and Pawlowski, Jan M. and Reichert, Manuel",
    title = "{Reconstructing the graviton}",
    eprint = "2102.02217",
    archivePrefix = "arXiv",
    primaryClass = "hep-th",
    doi = "10.21468/SciPostPhys.12.1.001",
    journal = "SciPost Phys.",
    volume = "12",
    number = "1",
    pages = "001",
    year = "2022"
}

@article{Eichhorn:2019ybe,
    author = "Eichhorn, Astrid and Platania, Alessia and Schiffer, Marc",
    title = "{Lorentz invariance violations in the interplay of quantum gravity with matter}",
    eprint = "1911.10066",
    archivePrefix = "arXiv",
    primaryClass = "hep-th",
    doi = "10.1103/PhysRevD.102.026007",
    journal = "Phys. Rev. D",
    volume = "102",
    number = "2",
    pages = "026007",
    year = "2020"
}

@inbook{Knorr:2022dsx,
    author = "Knorr, Benjamin and Ripken, Chris and Saueressig, Frank",
    title = "{Form Factors in Asymptotically Safe Quantum Gravity}",
    eprint = "2210.16072",
    archivePrefix = "arXiv",
    primaryClass = "hep-th",
    reportNumber = "NORDITA 2022-075",
    doi = "10.1007/978-981-19-3079-9_21-1",
    bookTitle="Handbook of Quantum Gravity",
    year="2023",
    publisher="Springer Nature Singapore",
    address="Singapore",
    pages="1--49"
}

@inbook{Morris:2022btf,
    author = "Morris, Tim R. and Stulga, Dalius",
    title = "{The Functional f(R) Approximation}",
    eprint = "2210.11356",
    archivePrefix = "arXiv",
    primaryClass = "hep-th",
    doi = "10.1007/978-981-19-3079-9_19-1",
    bookTitle="Handbook of Quantum Gravity",
    year="2023",
    publisher="Springer Nature Singapore",
    address="Singapore",
    pages="1--33"
}

@inbook{Martini:2022sll,
    author = "Martini, Riccardo and Vacca, Gian Paolo and Zanusso, Omar",
    title = "{Perturbative Approaches to Nonperturbative Quantum Gravity}",
    booktitle = "{Handbook of Quantum Gravity}",
    eprint = "2210.13910",
    archivePrefix = "arXiv",
    primaryClass = "hep-th",
    doi = "10.1007/978-981-19-3079-9_25-1",
    year="2023",
    publisher="Springer Nature Singapore",
    address="Singapore",
    pages="1--46"
}

@inbook{Wetterich:2022ncl,
    author = "Wetterich, C.",
    title = "{Quantum Gravity and Scale Symmetry in Cosmology}",
    eprint = "2211.03596",
    archivePrefix = "arXiv",
    primaryClass = "gr-qc",
    doi = "10.1007/978-981-19-3079-9_26-1",
    bookTitle="Handbook of Quantum Gravity",
    year="2023",
    publisher="Springer Nature Singapore",
    address="Singapore",
    pages="1--68"
}

@inbook{Saueressig:2023irs,
    author = "Saueressig, Frank",
    title = "{The Functional Renormalization Group in Quantum Gravity}",
    eprint = "2302.14152",
    archivePrefix = "arXiv",
    primaryClass = "hep-th",
    doi = "10.1007/978-981-19-3079-9_16-1",
    bookTitle="Handbook of Quantum Gravity",
    year="2023",
    publisher="Springer Nature Singapore",
    address="Singapore",
    pages="1--44"
}

@inbook{Pawlowski:2023gym,
    author = "Pawlowski, Jan M. and Reichert, Manuel",
    title = "{Quantum Gravity from Dynamical Metric Fluctuations}",
    eprint = "2309.10785",
    archivePrefix = "arXiv",
    primaryClass = "hep-th",
    bookTitle="Handbook of Quantum Gravity",
    year="2023",
    publisher="Springer Nature Singapore",
    address="Singapore",
    pages="1--70"
}

@inbook{Bonanno:2024xne,
    author = "Bonanno, Alfio",
    title = "{Asymptotic Safety and Cosmology}",
    doi = "10.1007/978-981-19-3079-9_23-1",
    bookTitle="Handbook of Quantum Gravity",
    year="2023",
    publisher="Springer Nature Singapore",
    address="Singapore",
    pages="1--27"
}

@article{Chen:2024sgx,
    author = "Chen, Calvin Y. -R. and de Rham, Claudia and Tolley, Andrew J.",
    title = "{Deformations of extremal black holes and the UV}",
    eprint = "2408.05549",
    archivePrefix = "arXiv",
    primaryClass = "hep-th",
    reportNumber = "Imperial/TP/2024/CC/1",
    doi = "10.1103/PhysRevD.111.024056",
    journal = "Phys. Rev. D",
    volume = "111",
    number = "2",
    pages = "024056",
    year = "2025"
}

@article{Shomer:2007vq,
    author = "Shomer, Assaf",
    title = "{A Pedagogical explanation for the non-renormalizability of gravity}",
    eprint = "0709.3555",
    archivePrefix = "arXiv",
    primaryClass = "hep-th",
    month = "9",
    year = "2007"
}

@article{Alberte:2020jsk,
    author = "Alberte, Lasma and de Rham, Claudia and Jaitly, Sumer and Tolley, Andrew J.",
    title = "{Positivity Bounds and the Massless Spin-2 Pole}",
    eprint = "2007.12667",
    archivePrefix = "arXiv",
    primaryClass = "hep-th",
    reportNumber = "Imperial/TP/2020/LA/02",
    doi = "10.1103/PhysRevD.102.125023",
    journal = "Phys. Rev. D",
    volume = "102",
    number = "12",
    pages = "125023",
    year = "2020"
}

@article{Basile:2023blg,
    author = {Basile, Ivano and L\"ust, Dieter and Montella, Carmine},
    title = "{Shedding black hole light on the emergent string conjecture}",
    eprint = "2311.12113",
    archivePrefix = "arXiv",
    primaryClass = "hep-th",
    reportNumber = "LMU-ASC 35/23, MPP-2023-262",
    doi = "10.1007/JHEP07(2024)208",
    journal = "JHEP",
    volume = "07",
    pages = "208",
    year = "2024"
}

@article{Castellano:2023aum,
    author = "Castellano, Alberto and Herr{\'a}ez, Alvaro and Ib{\'a}{\~n}ez, Luis E.",
    title = "{On the species scale, modular invariance and the gravitational EFT expansion}",
    eprint = "2310.07708",
    archivePrefix = "arXiv",
    primaryClass = "hep-th",
    doi = "10.1007/JHEP12(2024)019",
    journal = "JHEP",
    volume = "12",
    pages = "019",
    year = "2024"
}

@article{Basile:2024dqq,
    author = "Basile, Ivano and Cribiori, Niccol{\`o} and Lust, Dieter and Montella, Carmine",
    title = "{Minimal black holes and species thermodynamics}",
    eprint = "2401.06851",
    archivePrefix = "arXiv",
    primaryClass = "hep-th",
    reportNumber = "LMU-ASC 02/24, MPP-2024-6",
    doi = "10.1007/JHEP06(2024)127",
    journal = "JHEP",
    volume = "06",
    pages = "127",
    year = "2024"
}

@article{Bedroya:2024ubj,
    author = "Bedroya, Alek and Mishra, Rashmish K. and Wiesner, Max",
    title = "{Density of states, black holes and the Emergent String Conjecture}",
    eprint = "2405.00083",
    archivePrefix = "arXiv",
    primaryClass = "hep-th",
    doi = "10.1007/JHEP01(2025)144",
    journal = "JHEP",
    volume = "01",
    pages = "144",
    year = "2025"
}

@article{Aoufia:2024awo,
    author = "Aoufia, Christian and Basile, Ivano and Leone, Giorgio",
    title = "{Species scale, worldsheet CFTs and emergent geometry}",
    eprint = "2405.03683",
    archivePrefix = "arXiv",
    primaryClass = "hep-th",
    doi = "10.1007/JHEP12(2024)111",
    journal = "JHEP",
    volume = "12",
    pages = "111",
    year = "2024"
}

@article{Henriksson:2022oeu,
    author = "Henriksson, Johan and McPeak, Brian and Russo, Francesco and Vichi, Alessandro",
    title = "{Bounding violations of the weak gravity conjecture}",
    eprint = "2203.08164",
    archivePrefix = "arXiv",
    primaryClass = "hep-th",
    doi = "10.1007/JHEP08(2022)184",
    journal = "JHEP",
    volume = "08",
    pages = "184",
    year = "2022"
}

@article{Herrero-Valea:2020wxz,
    author = "Herrero-Valea, Mario and Santos-Garcia, Raquel and Tokareva, Anna",
    title = "{Massless positivity in graviton exchange}",
    eprint = "2011.11652",
    archivePrefix = "arXiv",
    primaryClass = "hep-th",
    reportNumber = "FTUAM-20-26, IFT-UAM/CSIC-20-162, INR-TH-2020-043",
    doi = "10.1103/PhysRevD.104.085022",
    journal = "Phys. Rev. D",
    volume = "104",
    number = "8",
    pages = "085022",
    year = "2021"
}

@article{Herrero-Valea:2022lfd,
    author = "Herrero-Valea, M. and Koshelev, A. S. and Tokareva, A.",
    title = "{UV graviton scattering and positivity bounds from IR dispersion relations}",
    eprint = "2205.13332",
    archivePrefix = "arXiv",
    primaryClass = "hep-th",
    reportNumber = "Imperial/TP/2022/AAT/1",
    doi = "10.1103/PhysRevD.106.105002",
    journal = "Phys. Rev. D",
    volume = "106",
    number = "10",
    pages = "105002",
    year = "2022"
}

@book{Bambi:2023jiz,
    editor = "Bambi, Cosimo and Modesto, Leonardo and Shapiro, Ilya",
    title = "{Handbook of Quantum Gravity}",
    doi = "10.1007/978-981-99-7681-2",
    isbn = "978-981-99-7680-5, 978-981-99-7681-2, 978-981-19-3079-9",
    publisher = "Springer",
    year = "2024"
}

@article{Eichhorn:2025sux,
    author = "Eichhorn, Astrid and Gyftopoulos, Zois and Held, Aaron",
    title = "{Quark and lepton mixing in the asymptotically safe Standard Model}",
    eprint = "2507.18304",
    archivePrefix = "arXiv",
    primaryClass = "hep-ph",
    month = "7",
    year = "2025"
}

@article{Gies:2015tca,
    author = "Gies, Holger and Knorr, Benjamin and Lippoldt, Stefan",
    title = "{Generalized Parametrization Dependence in Quantum Gravity}",
    eprint = "1507.08859",
    archivePrefix = "arXiv",
    primaryClass = "hep-th",
    doi = "10.1103/PhysRevD.92.084020",
    journal = "Phys. Rev. D",
    volume = "92",
    number = "8",
    pages = "084020",
    year = "2015"
}

@article{Knorr:2023usb,
    author = "Knorr, Benjamin",
    title = "{Momentum-dependent field redefinitions in asymptotic safety}",
    eprint = "2311.12097",
    archivePrefix = "arXiv",
    primaryClass = "hep-th",
    reportNumber = "NORDITA 2023-065",
    doi = "10.1103/PhysRevD.110.026001",
    journal = "Phys. Rev. D",
    volume = "110",
    number = "2",
    pages = "026001",
    year = "2024"
}

@article{Draper:2020knh,
    author = "Draper, Tom and Knorr, Benjamin and Ripken, Chris and Saueressig, Frank",
    title = "{Graviton-Mediated Scattering Amplitudes from the Quantum Effective Action}",
    eprint = "2007.04396",
    archivePrefix = "arXiv",
    primaryClass = "hep-th",
    doi = "10.1007/JHEP11(2020)136",
    journal = "JHEP",
    volume = "11",
    pages = "136",
    year = "2020"
}

@article{Buccio:2023lzo,
    author = "Buccio, Diego and Donoghue, John F. and Percacci, Roberto",
    title = "{Amplitudes and renormalization group techniques: A case study}",
    eprint = "2307.00055",
    archivePrefix = "arXiv",
    primaryClass = "hep-th",
    doi = "10.1103/PhysRevD.109.045008",
    journal = "Phys. Rev. D",
    volume = "109",
    number = "4",
    pages = "045008",
    year = "2024"
}

@article{deAlwis:2019aud,
    author = "de Alwis, Senarath and Eichhorn, Astrid and Held, Aaron and Pawlowski, Jan M. and Schiffer, Marc and Versteegen, Fleur",
    title = "{Asymptotic safety, string theory and the weak gravity conjecture}",
    eprint = "1907.07894",
    archivePrefix = "arXiv",
    primaryClass = "hep-th",
    doi = "10.1016/j.physletb.2019.134991",
    journal = "Phys. Lett. B",
    volume = "798",
    pages = "134991",
    year = "2019"
}

@book{Percacci:2017fkn,
    author = "Percacci, Robert",
    title = "{An Introduction to Covariant Quantum Gravity and Asymptotic Safety}",
    doi = "10.1142/10369",
    isbn = "978-981-320-717-2, 978-981-320-719-6",
    publisher = "World Scientific",
    series = "100 Years of General Relativity",
    volume = "3",
    year = "2017"
}

@article{Knorr:2021iwv,
    author = "Knorr, Benjamin and Ripken, Chris and Saueressig, Frank",
    title = "{Form Factors in Quantum Gravity: Contrasting non-local, ghost-free gravity and Asymptotic Safety}",
    eprint = "2111.12365",
    archivePrefix = "arXiv",
    primaryClass = "hep-th",
    doi = "10.1393/ncc/i2022-22028-5",
    journal = "Nuovo Cim. C",
    volume = "45",
    number = "2",
    pages = "28",
    year = "2022"
}

@article{Dupuis:2020fhh,
    author = "Dupuis, N. and Canet, L. and Eichhorn, A. and Metzner, W. and Pawlowski, J. M. and Tissier, M. and Wschebor, N.",
    title = "{The nonperturbative functional renormalization group and its applications}",
    eprint = "2006.04853",
    archivePrefix = "arXiv",
    primaryClass = "cond-mat.stat-mech",
    doi = "10.1016/j.physrep.2021.01.001",
    journal = "Phys. Rept.",
    volume = "910",
    pages = "1--114",
    year = "2021"
}

@inbook{Platania:2023srt,
    author = "Platania, Alessia",
    title = "{Black Holes in Asymptotically Safe Gravity}",
    eprint = "2302.04272",
    archivePrefix = "arXiv",
    primaryClass = "gr-qc",
    reportNumber = "NORDITA 2022-085",
    doi = "10.1007/978-981-19-3079-9_24-1",
    month = "2",
    bookTitle="Handbook of Quantum Gravity",
    publisher="Springer Nature Singapore",
    address="Singapore",
    year = "2023",
    pages="1--65"
}

@article{Reuter:1996cp,
    author = "Reuter, M.",
    title = "{Nonperturbative evolution equation for quantum gravity}",
    eprint = "hep-th/9605030",
    archivePrefix = "arXiv",
    reportNumber = "DESY-96-080",
    doi = "10.1103/PhysRevD.57.971",
    journal = "Phys. Rev. D",
    volume = "57",
    pages = "971--985",
    year = "1998"
}

@article{Bonanno:2020bil,
    author = "Bonanno, Alfio and Eichhorn, Astrid and Gies, Holger and Pawlowski, Jan M. and Percacci, Roberto and Reuter, Martin and Saueressig, Frank and Vacca, Gian Paolo",
    title = "{Critical reflections on asymptotically safe gravity}",
    eprint = "2004.06810",
    archivePrefix = "arXiv",
    primaryClass = "gr-qc",
    doi = "10.3389/fphy.2020.00269",
    journal = "Front. in Phys.",
    volume = "8",
    pages = "269",
    year = "2020"
}

@article{Pastor-Gutierrez:2022nki,
    author = "Pastor-Guti{\'e}rrez, {\'A}lvaro and Pawlowski, Jan M. and Reichert, Manuel",
    title = "{The Asymptotically Safe Standard Model: From quantum gravity to dynamical chiral symmetry breaking}",
    eprint = "2207.09817",
    archivePrefix = "arXiv",
    primaryClass = "hep-th",
    doi = "10.21468/SciPostPhys.15.3.105",
    journal = "SciPost Phys.",
    volume = "15",
    number = "3",
    pages = "105",
    year = "2023"
}

@article{Knorr:2021slg,
    author = "Knorr, Benjamin",
    title = "{The derivative expansion in asymptotically safe quantum gravity: general setup and quartic order}",
    eprint = "2104.11336",
    archivePrefix = "arXiv",
    primaryClass = "hep-th",
    doi = "10.21468/SciPostPhysCore.4.3.020",
    journal = "SciPost Phys. Core",
    volume = "4",
    pages = "020",
    year = "2021"
}

@article{Gies:2016con,
    author = "Gies, Holger and Knorr, Benjamin and Lippoldt, Stefan and Saueressig, Frank",
    title = "{Gravitational Two-Loop Counterterm Is Asymptotically Safe}",
    eprint = "1601.01800",
    archivePrefix = "arXiv",
    primaryClass = "hep-th",
    doi = "10.1103/PhysRevLett.116.211302",
    journal = "Phys. Rev. Lett.",
    volume = "116",
    number = "21",
    pages = "211302",
    year = "2016"
}

@article{Knorr:2019atm,
    author = "Knorr, Benjamin and Ripken, Chris and Saueressig, Frank",
    title = "{Form Factors in Asymptotic Safety: conceptual ideas and computational toolbox}",
    eprint = "1907.02903",
    archivePrefix = "arXiv",
    primaryClass = "hep-th",
    doi = "10.1088/1361-6382/ab4a53",
    journal = "Class. Quant. Grav.",
    volume = "36",
    number = "23",
    pages = "234001",
    year = "2019"
}

@article{Baldazzi:2021ydj,
    author = "Baldazzi, Alessio and Zinati, Riccardo Ben Al{\`\i} and Falls, Kevin",
    title = "{Essential renormalisation group}",
    eprint = "2105.11482",
    archivePrefix = "arXiv",
    primaryClass = "hep-th",
    doi = "10.21468/SciPostPhys.13.4.085",
    journal = "SciPost Phys.",
    volume = "13",
    number = "4",
    pages = "085",
    year = "2022"
}

@article{Baldazzi:2021orb,
    author = "Baldazzi, Alessio and Falls, Kevin",
    title = "{Essential Quantum Einstein Gravity}",
    eprint = "2107.00671",
    archivePrefix = "arXiv",
    primaryClass = "hep-th",
    doi = "10.3390/universe7080294",
    journal = "Universe",
    volume = "7",
    number = "8",
    pages = "294",
    year = "2021"
}

@article{Basile:2021euh,
    author = "Basile, Ivano and Platania, Alessia",
    title = "{Cosmological \ensuremath{\alpha}'-corrections from the functional renormalization group}",
    eprint = "2101.02226",
    archivePrefix = "arXiv",
    primaryClass = "hep-th",
    doi = "10.1007/JHEP06(2021)045",
    journal = "JHEP",
    volume = "06",
    pages = "045",
    year = "2021"
}

@article{Basile:2021krk,
    author = "Basile, Ivano and Platania, Alessia",
    title = "{String tension between de Sitter vacua and curvature corrections}",
    eprint = "2103.06276",
    archivePrefix = "arXiv",
    primaryClass = "hep-th",
    doi = "10.1103/PhysRevD.104.L121901",
    journal = "Phys. Rev. D",
    volume = "104",
    number = "12",
    pages = "L121901",
    year = "2021"
}

@article{Basile:2021krr,
    author = "Basile, Ivano and Platania, Alessia",
    title = "{Asymptotic Safety: Swampland or Wonderland?}",
    eprint = "2107.06897",
    archivePrefix = "arXiv",
    primaryClass = "hep-th",
    doi = "10.3390/universe7100389",
    journal = "Universe",
    volume = "7",
    number = "10",
    pages = "389",
    year = "2021"
}

@article{Horowitz:2023xyl,
    author = "Horowitz, Gary T. and Kolanowski, Maciej and Remmen, Grant N. and Santos, Jorge E.",
    title = "{Extremal Kerr Black Holes as Amplifiers of New Physics}",
    eprint = "2303.07358",
    archivePrefix = "arXiv",
    primaryClass = "hep-th",
    doi = "10.1103/PhysRevLett.131.091402",
    journal = "Phys. Rev. Lett.",
    volume = "131",
    number = "9",
    pages = "091402",
    year = "2023"}

@article{Goroff:1985sz,
    author = "Goroff, Marc H. and Sagnotti, Augusto",
    title = "{QUANTUM GRAVITY AT TWO LOOPS}",
    reportNumber = "CALT-68-1263, UCB-PTH-85/18, LBL-19512",
    doi = "10.1016/0370-2693(85)91470-4",
    journal = "Phys. Lett. B",
    volume = "160",
    pages = "81--86",
    year = "1985"
}

@article{Goroff:1985th,
    author = "Goroff, Marc H. and Sagnotti, Augusto",
    title = "{The Ultraviolet Behavior of Einstein Gravity}",
    reportNumber = "CALT-68-1289, LBL-19995, UCB-PTH-85-34",
    doi = "10.1016/0550-3213(86)90193-8",
    journal = "Nucl. Phys. B",
    volume = "266",
    pages = "709--736",
    year = "1986"
}

@article{vandeVen:1991gw,
    author = "van de Ven, A. E. M.",
    title = "{Two loop quantum gravity}",
    reportNumber = "DESY-91-115, ITP-SB-91-52",
    doi = "10.1016/0550-3213(92)90011-Y",
    journal = "Nucl. Phys. B",
    volume = "378",
    pages = "309--366",
    year = "1992"
}

@article{Baldazzi:2023pep,
    author = "Baldazzi, Alessio and Falls, Kevin and Kluth, Yannick and Knorr, Benjamin",
    title = "{Robustness of the derivative expansion in Asymptotic Safety}",
    eprint = "2312.03831",
    archivePrefix = "arXiv",
    primaryClass = "hep-th",
    reportNumber = "NORDITA 2023-075",
    month = "12",
    year = "2023"
}

@article{Basile:2025zjc,
    author = "Basile, Ivano and Knorr, Benjamin and Platania, Alessia and Schiffer, Marc",
    title = "{Asymptotic safety, quantum gravity, and the swampland: a conceptual assessment}",
    eprint = "2502.12290",
    archivePrefix = "arXiv",
    primaryClass = "hep-th",
    reportNumber = "MPP-2025-37",
    month = "2",
    year = "2025"
}

@article{Knorr:2024yiu,
    author = "Knorr, Benjamin and Platania, Alessia",
    title = "{Unearthing the intersections: positivity bounds, weak gravity conjecture, and asymptotic safety landscapes from photon-graviton flows}",
    eprint = "2405.08860",
    archivePrefix = "arXiv",
    primaryClass = "hep-th",
    reportNumber = "NORDITA 2024-014",
    doi = "10.1007/JHEP03(2025)003",
    journal = "JHEP",
    volume = "03",
    pages = "003",
    year = "2025"
}

@article{Horowitz:2024dch,
    author = "Horowitz, Gary T. and Kolanowski, Maciej and Remmen, Grant N. and Santos, Jorge E.",
    title = "{Sudden breakdown of effective field theory near cool Kerr-Newman black holes}",
    eprint = "2403.00051",
    archivePrefix = "arXiv",
    primaryClass = "hep-th",
    doi = "10.1007/JHEP05(2024)122",
    journal = "JHEP",
    volume = "05",
    pages = "122",
    year = "2024"
}

@article{Arkani-Hamed:2006emk,
    author = "Arkani-Hamed, Nima and Motl, Lubos and Nicolis, Alberto and Vafa, Cumrun",
    title = "{The String landscape, black holes and gravity as the weakest force}",
    eprint = "hep-th/0601001",
    archivePrefix = "arXiv",
    reportNumber = "HUTP-05-A0057",
    doi = "10.1088/1126-6708/2007/06/060",
    journal = "JHEP",
    volume = "06",
    pages = "060",
    year = "2007"
}

@article{Cheung:2018cwt,
    author = "Cheung, Clifford and Liu, Junyu and Remmen, Grant N.",
    title = "{Proof of the Weak Gravity Conjecture from Black Hole Entropy}",
    eprint = "1801.08546",
    archivePrefix = "arXiv",
    primaryClass = "hep-th",
    reportNumber = "CALT-TH-2018-007",
    doi = "10.1007/JHEP10(2018)004",
    journal = "JHEP",
    volume = "10",
    pages = "004",
    year = "2018"
}

@article{Hamada:2018dde,
    author = "Hamada, Yuta and Noumi, Toshifumi and Shiu, Gary",
    title = "{Weak Gravity Conjecture from Unitarity and Causality}",
    eprint = "1810.03637",
    archivePrefix = "arXiv",
    primaryClass = "hep-th",
    reportNumber = "CCTP-2018-12, ITCP-IPP 2018/9, KOBE-COSMO-18-08, MAD-TH-18-05",
    doi = "10.1103/PhysRevLett.123.051601",
    journal = "Phys. Rev. Lett.",
    volume = "123",
    number = "5",
    pages = "051601",
    year = "2019"
}

@article{Urbano:2018kax,
    author = "Urbano, Alfredo",
    title = "{Towards a proof of the Weak Gravity Conjecture}",
    eprint = "1810.05621",
    archivePrefix = "arXiv",
    primaryClass = "hep-th",
    month = "10",
    year = "2018"
}

@article{Banks:2006mm,
    author = "Banks, Tom and Johnson, Matt and Shomer, Assaf",
    title = "{A Note on Gauge Theories Coupled to Gravity}",
    eprint = "hep-th/0606277",
    archivePrefix = "arXiv",
    reportNumber = "RU-06-08, SCIPP-06-07",
    doi = "10.1088/1126-6708/2006/09/049",
    journal = "JHEP",
    volume = "09",
    pages = "049",
    year = "2006"
}

@article{Heidenreich:2024dmr,
    author = "Heidenreich, Ben and Lotito, Matteo",
    title = "{Proving the Weak Gravity Conjecture in perturbative string theory. Part I. The bosonic string}",
    eprint = "2401.14449",
    archivePrefix = "arXiv",
    primaryClass = "hep-th",
    reportNumber = "ACF-T24-01, ACFI-T24-01",
    doi = "10.1007/JHEP05(2025)102",
    journal = "JHEP",
    volume = "05",
    pages = "102",
    year = "2025"
}

@article{Falls:2020qhj,
    author = "Falls, Kevin and Ohta, Nobuyoshi and Percacci, Roberto",
    title = "{Towards the determination of the dimension of the critical surface in asymptotically safe gravity}",
    eprint = "2004.04126",
    archivePrefix = "arXiv",
    primaryClass = "hep-th",
    doi = "10.1016/j.physletb.2020.135773",
    journal = "Phys. Lett. B",
    volume = "810",
    pages = "135773",
    year = "2020"
}

@article{Kluth:2020bdv,
    author = "Kluth, Yannick and Litim, Daniel F.",
    title = "{Fixed points of quantum gravity and the dimensionality of the UV critical surface}",
    eprint = "2008.09181",
    archivePrefix = "arXiv",
    primaryClass = "hep-th",
    doi = "10.1103/PhysRevD.108.026005",
    journal = "Phys. Rev. D",
    volume = "108",
    number = "2",
    pages = "026005",
    year = "2023"
}

@article{Platania:2025imw,
    author = "Platania, Alessia",
    title = "{Some thoughts about black holes in asymptotic safety}",
    doi = "10.1007/s10714-025-03390-5",
    journal = "Gen. Rel. Grav.",
    volume = "57",
    number = "3",
    pages = "58",
    year = "2025"
}

@article{Shaposhnikov:2009pv,
    author = "Shaposhnikov, Mikhail and Wetterich, Christof",
    title = "{Asymptotic safety of gravity and the Higgs boson mass}",
    eprint = "0912.0208",
    archivePrefix = "arXiv",
    primaryClass = "hep-th",
    doi = "10.1016/j.physletb.2009.12.022",
    journal = "Phys. Lett. B",
    volume = "683",
    pages = "196--200",
    year = "2010"
}

@article{Eichhorn:2017lry,
    author = "Eichhorn, Astrid and Versteegen, Fleur",
    title = "{Upper bound on the Abelian gauge coupling from asymptotic safety}",
    eprint = "1709.07252",
    archivePrefix = "arXiv",
    primaryClass = "hep-th",
    doi = "10.1007/JHEP01(2018)030",
    journal = "JHEP",
    volume = "01",
    pages = "030",
    year = "2018"
}

@article{Eichhorn:2018whv,
    author = "Eichhorn, Astrid and Held, Aaron",
    title = "{Mass difference for charged quarks from asymptotically safe quantum gravity}",
    eprint = "1803.04027",
    archivePrefix = "arXiv",
    primaryClass = "hep-th",
    doi = "10.1103/PhysRevLett.121.151302",
    journal = "Phys. Rev. Lett.",
    volume = "121",
    number = "15",
    pages = "151302",
    year = "2018"
}

@article{Eichhorn:2017ylw,
    author = "Eichhorn, Astrid and Held, Aaron",
    title = "{Top mass from asymptotic safety}",
    eprint = "1707.01107",
    archivePrefix = "arXiv",
    primaryClass = "hep-th",
    doi = "10.1016/j.physletb.2017.12.040",
    journal = "Phys. Lett. B",
    volume = "777",
    pages = "217--221",
    year = "2018"
}

@article{Percacci:2010af,
    author = "Percacci, Roberto and Vacca, Gian Paolo",
    title = "{Asymptotic Safety, Emergence and Minimal Length}",
    eprint = "1008.3621",
    archivePrefix = "arXiv",
    primaryClass = "hep-th",
    doi = "10.1088/0264-9381/27/24/245026",
    journal = "Class. Quant. Grav.",
    volume = "27",
    pages = "245026",
    year = "2010"
}

@article{Held:2020kze,
    author = "Held, Aaron",
    title = "{Effective asymptotic safety and its predictive power: Gauge-Yukawa theories}",
    eprint = "2003.13642",
    archivePrefix = "arXiv",
    primaryClass = "hep-th",
    reportNumber = "Imperial/TP/2020/AH/02",
    doi = "10.3389/fphy.2020.00341",
    journal = "Front. in Phys.",
    volume = "8",
    pages = "341",
    year = "2020"
}

@article{Reuter:2001ag,
    author = "Reuter, M. and Saueressig, Frank",
    title = "{Renormalization group flow of quantum gravity in the Einstein-Hilbert truncation}",
    eprint = "hep-th/0110054",
    archivePrefix = "arXiv",
    reportNumber = "MZ-TH-01-27",
    doi = "10.1103/PhysRevD.65.065016",
    journal = "Phys. Rev. D",
    volume = "65",
    pages = "065016",
    year = "2002"
}

@article{Codello:2006in,
    author = "Codello, Alessandro and Percacci, Roberto",
    title = "{Fixed points of higher derivative gravity}",
    eprint = "hep-th/0607128",
    archivePrefix = "arXiv",
    doi = "10.1103/PhysRevLett.97.221301",
    journal = "Phys. Rev. Lett.",
    volume = "97",
    pages = "221301",
    year = "2006"
}

@article{Lauscher:2002sq,
    author = "Lauscher, O. and Reuter, M.",
    title = "{Flow equation of quantum Einstein gravity in a higher derivative truncation}",
    eprint = "hep-th/0205062",
    archivePrefix = "arXiv",
    reportNumber = "MZ-TH-02-07",
    doi = "10.1103/PhysRevD.66.025026",
    journal = "Phys. Rev. D",
    volume = "66",
    pages = "025026",
    year = "2002"
}

@article{Hamada:2017rvn,
    author = "Hamada, Yuta and Yamada, Masatoshi",
    title = "{Asymptotic safety of higher derivative quantum gravity non-minimally coupled with a matter system}",
    eprint = "1703.09033",
    archivePrefix = "arXiv",
    primaryClass = "hep-th",
    doi = "10.1007/JHEP08(2017)070",
    journal = "JHEP",
    volume = "08",
    pages = "070",
    year = "2017"
}

@article{Falls:2017lst,
    author = "Falls, Kevin and King, Callum R. and Litim, Daniel F. and Nikolakopoulos, Kostas and Rahmede, Christoph",
    title = "{Asymptotic safety of quantum gravity beyond Ricci scalars}",
    eprint = "1801.00162",
    archivePrefix = "arXiv",
    primaryClass = "hep-th",
    doi = "10.1103/PhysRevD.97.086006",
    journal = "Phys. Rev. D",
    volume = "97",
    number = "8",
    pages = "086006",
    year = "2018"
}

@article{Fulling:1992vm,
    author = "Fulling, S. A. and King, Ronald C. and Wybourne, B. G. and Cummins, C. J.",
    title = "{Normal forms for tensor polynomials. 1: The Riemann tensor}",
    doi = "10.1088/0264-9381/9/5/003",
    journal = "Class. Quant. Grav.",
    volume = "9",
    pages = "1151--1197",
    year = "1992"
}

@article{Harlow:2022ich,
    author = "Harlow, Daniel and Heidenreich, Ben and Reece, Matthew and Rudelius, Tom",
    title = "{Weak gravity conjecture}",
    eprint = "2201.08380",
    archivePrefix = "arXiv",
    primaryClass = "hep-th",
    reportNumber = "ACFI-T22-01",
    doi = "10.1103/RevModPhys.95.035003",
    journal = "Rev. Mod. Phys.",
    volume = "95",
    number = "3",
    pages = "035003",
    year = "2023"
}

@article{Buccio:2024hys,
    author = "Buccio, Diego and Donoghue, John F. and Menezes, Gabriel and Percacci, Roberto",
    title = "{Physical Running of Couplings in Quadratic Gravity}",
    eprint = "2403.02397",
    archivePrefix = "arXiv",
    primaryClass = "hep-th",
    doi = "10.1103/PhysRevLett.133.021604",
    journal = "Phys. Rev. Lett.",
    volume = "133",
    number = "2",
    pages = "021604",
    year = "2024"
}

@article{Tokareva:2025rta,
    author = "Tokareva, Anna and Xu, Yongjun",
    title = "{Scalar weak gravity bound from full unitarity}",
    eprint = "2502.10375",
    archivePrefix = "arXiv",
    primaryClass = "hep-th",
    month = "2",
    year = "2025"
}

@inbook{Eichhorn:2022gku,
    author = "Eichhorn, Astrid and Schiffer, Marc",
    title = "Asymptotic Safety of Gravity with Matter",
    eprint = "2212.07456",
    archivePrefix = "arXiv",
    primaryClass = "hep-th",
    bookTitle="Handbook of Quantum Gravity",
    year="2023",
    publisher="Springer Nature Singapore",
    address="Singapore",
    pages="1--87"
}

@Article{Basile:2024oms,
	title={{Lectures in quantum gravity}},
	author={Ivano Basile and Luca Buoninfante and Francesco Di Filippo and Benjamin Knorr and Alessia Platania and Anna Tokareva},
	journal={SciPost Phys. Lect. Notes},
	pages={98},
	year={2025},
	publisher={SciPost},
	doi={10.21468/SciPostPhysLectNotes.98},
    eprint = "2412.08690",
    archivePrefix = "arXiv",
    primaryClass = "hep-th",
	url={https://scipost.org/10.21468/SciPostPhysLectNotes.98},
}

@article{Buoninfante:2024yth,
    author = "Buoninfante, Luca and others",
    title = "{Visions in Quantum Gravity}",
    eprint = "2412.08696",
    archivePrefix = "arXiv",
    primaryClass = "hep-th",
    journal={SciPost Phys. Comm. Rep.},
	pages={11},
	year={2025},
	publisher={SciPost},
	doi={10.21468/SciPostPhysCommRep.11}
}

@article{Dona:2013qba,
    author = "Don{\`a}, Pietro and Eichhorn, Astrid and Percacci, Roberto",
    title = "{Matter matters in asymptotically safe quantum gravity}",
    eprint = "1311.2898",
    archivePrefix = "arXiv",
    primaryClass = "hep-th",
    doi = "10.1103/PhysRevD.89.084035",
    journal = "Phys. Rev. D",
    volume = "89",
    number = "8",
    pages = "084035",
    year = "2014"
}

@article{Denz:2016qks,
    author = "Denz, Tobias and Pawlowski, Jan M. and Reichert, Manuel",
    title = "{Towards apparent convergence in asymptotically safe quantum gravity}",
    eprint = "1612.07315",
    archivePrefix = "arXiv",
    primaryClass = "hep-th",
    doi = "10.1140/epjc/s10052-018-5806-0",
    journal = "Eur. Phys. J. C",
    volume = "78",
    number = "4",
    pages = "336",
    year = "2018"
}

@article{Donoghue:2015xla,
    author = "Donoghue, John F. and El-Menoufi, Basem Kamal",
    title = "{QED trace anomaly, non-local Lagrangians and quantum Equivalence Principle violations}",
    eprint = "1503.06099",
    archivePrefix = "arXiv",
    primaryClass = "hep-th",
    doi = "10.1007/JHEP05(2015)118",
    journal = "JHEP",
    volume = "05",
    pages = "118",
    year = "2015"
}

@article{Donoghue:2018izj,
    author = "Donoghue, John F. and Menezes, Gabriel",
    title = "{Gauge Assisted Quadratic Gravity: A Framework for UV Complete Quantum Gravity}",
    eprint = "1804.04980",
    archivePrefix = "arXiv",
    primaryClass = "hep-th",
    reportNumber = "ACFI- T18-06, ACFI-T18-06, ACFI--T18-06",
    doi = "10.1103/PhysRevD.97.126005",
    journal = "Phys. Rev. D",
    volume = "97",
    number = "12",
    pages = "126005",
    year = "2018"
}

@article{Donoghue:2019fcb,
    author = "Donoghue, John F. and Menezes, Gabriel",
    title = "{Unitarity, stability and loops of unstable ghosts}",
    eprint = "1908.02416",
    archivePrefix = "arXiv",
    primaryClass = "hep-th",
    reportNumber = "ACFI-T19-08",
    doi = "10.1103/PhysRevD.100.105006",
    journal = "Phys. Rev. D",
    volume = "100",
    number = "10",
    pages = "105006",
    year = "2019"
}

\end{document}